\documentclass[12pt]{article}
\usepackage{amsmath,amssymb,amsfonts}
\usepackage[paper=letterpaper,margin=1.0in]{geometry}
\usepackage{graphicx}

\newcommand{\be}{\begin{equation}}
\newcommand{\ee}{\end{equation}}
\newcommand{\bea}{\begin{eqnarray}}
\newcommand{\eea}{\end{eqnarray}}
\newcommand{\ba}{\begin{array}}
\newcommand{\ea}{\end{array}}
\newcommand{\ben}{\begin{enumerate}}
\newcommand{\een}{\end{enumerate}}
\newcommand{\bi}{\begin{itemize}}
\newcommand{\ei}{\end{itemize}}
\newcommand{\bc}{\begin{center}}
\newcommand{\ec}{\end{center}}
\newcommand{\bfig}{\begin{figure}}
\newcommand{\efig}{\end{figure}}
\newcommand{\bq}{\begin{quotation}}
\newcommand{\eq}{\end{quotation}}
\newcommand{\bt}{\begin{table}}
\newcommand{\et}{\end{table}}
\newcommand{\btab}{\begin{tabular}}
\newcommand{\etab}{\end{tabular}}
\newcommand{\bs}{\begin{slide}}
\newcommand{\es}{\end{slide}}

\begin{document}

{\footnotesize
${}$
}

\bc

\vskip 1.0cm
\centerline{\Large \bf Quantum gravity \&  BH-NS binaries}
\vskip 0.5cm
\vskip 1.0cm

\renewcommand{\thefootnote}{\fnsymbol{footnote}}

\centerline{{\bf
Michael J. Kavic${}^{1}$\footnote{\tt michael.kavic@liu.edu},
Djordje Minic${}^{2}$\footnote{\tt dminic@vt.edu}
and
John Simonetti${}^{2}$\footnote{\tt jhs@vt.edu}
}}

\vskip 0.5cm

{\it
${}^1$Department of Physics, Long Island University, Brooklyn, New York 11201, U.S.A. \\
${}^2$Department  of Physics, Virginia Tech, Blacksburg, VA 24061, U.S.A. \\
${}$ \\
}

\ec

\vskip 1.0cm

\begin{abstract}
 We argue that the Black Hole-Neutron Star (BH-NS) binaries are the natural astrophysical probes of quantum gravity in the context of the new era of multi-messenger astronomy. In particular, we discuss the observable effect of enhanced black-hole mass loss in a BH-NS binary, due to the presence of an additional length scale tied to the intrinsic non-commutativity of quantum spacetime in quantum gravity.
\end{abstract}

\vspace{1cm}

\begin{center}

This essay received an honorable mention in the Gravity Research Foundation 2018 Awards for Essays on Gravitation competition.\\



\end{center}

{\it In memory of Joe Polchinski}

\renewcommand{\thefootnote}{\arabic{footnote}}

\newpage


{\bf Introduction:} 
The two great legacies of the 20th century physics are the frameworks of relativity (special and general) and quantum theory (including local quantum field theory). Both have been tested in numerous experiments, including the recent spectacular discovery of gravitational waves (GW) by LIGO \cite{ligo}, and the completion of the fundamental quantum field theoretic framework of the Standard Model of particle physics, exemplified by the LHC discovery of the Higgs particle \cite{higgs}. However, these two great frameworks have not been reconciled, and so we still live with the puzzle of quantum gravity.

{\bf The puzzle of quantum gravity:} 
The conceptual foundations of general relativity and quantum theory thus still stand apart.
It is often stated that one of the most outstanding problems of physics is to find a unified habitat for these two
great frameworks \cite{qg}. Apart from the obvious theoretical challenge, the observational/experimental challenge is even more outstanding.
 However, this claim might be ripe for a revision given the opening of the new field of gravitational wave astronomy \cite{ligo} and its multi-messenger counterparts. 
 
 As this new era in observational astrophysics dawns, new approaches to quantum gravity are coming to light which challenge long-held assumptions about non-locality and the scales at which quantum gravity is manifest.
The conventional claim is that quantum gravity is a Planck scale $l_P = \sqrt{\hbar G_N/c^3}$ effect, and thus, out of reach of present day experiments and observations. However, this claim is based on the assumptions of locality and separation of high energy and low energy scales. These assumptions have been tested in numerous experiments, but these tests do not preclude a possible fundamental mixing between short and long distance scales, and thus the possibility of quantum gravity phenomenology 
at observable scales.
In the past we have argued that astrophysics is the perfect place to test possible approaches to quantum gravity 
\cite{Kavic:2008wg, Simonetti:2010mk, Estes:2016wgv, Liebling:2017pqs}. In this essay we want to argue that Black Hole-Neutron Star (BH-NS) binaries are the natural astrophysical probes of quantum gravity in the context of the new era of multi-messenger astronomy and the new theoretical insights on the problem of quantum gravity.

{\bf Quantum gravity $=$ gravitizing the quantum:}
Recently a new insight on the vexing question of quantum gravity has been provided in the work of Freidel, Leigh and Minic \cite{flm}.
In what follows we base our discussion regarding the observable effects of quantum gravity on this work. According to \cite{flm}, quantum gravity can be viewed as a theory that reaches to the foundations of quantum theory in terms of a manifestly non-local formulation of quantum mechanics (including quantum field theory). This new approach is based on a quantum spacetime whose geometry (consisting of mutually compatible antisymmetric symplectic structure $\omega_{AB}$, the symmetric polarization metric $\eta_{AB}$ and the doubled symmetric metric $H_{AB}$) captures the essential quantum non-locality of any quantum theory. Quantum gravity becomes that quantum theory whose intrinsic geometry ($\omega_{AB}$, $\eta_{AB}$, $H_{AB}$) of quantum non-locality is made dynamical  \cite{flm}.
Thus the slogan, ``quantum gravity $=$ gravitizing the quantum.''

The work of \cite{flm} reformulates what is meant by quantum gravity in the guise of a new formulation of string theory. 
To begin with, this approach offers a precise notion of quantum spacetime, what quantum
superpositions should mean, and offers a clear path to the emergence of large scale classical spacetimes. 
In particular, the geometry of quantum theory in its manifestly non-local formulation is revealed, which leads to new insights involving noncommutativity in string theory and its deepest symmetries. According to \cite{flm}, quantum gravity should not be understood in terms of quantized
fields on classical spacetimes, but on quantizing spacetime itself. The approach presented in \cite{flm} is not ad hoc, but is built upon the familiar and quantum mechanically consistent confines of string theory itself.

Thus, \cite{flm} is a fundamentally new approach to quantum gravity that focuses on the foundations of space and time from the point of view of quantum theoretical probes. Traditionally \cite{qg}, quantum gravity is understood as a formulation of the problem of quantizing a collection of fields labeled by classical spacetime points and it is expected that its description can be recast in terms of effective field theory. What this conventional approach \cite{qg} implies is that the effective description of quantum gravity does not affect spacetime itself, just the physical perturbations, and it presupposes the presence of locality as an organizing principle. Although this approach is efficient for the leading quantum gravitational effects \cite{qg}, we do not expect it to be fundamental. The program of \cite{flm}, proposes to go deeper and revise the notion of locality itself by allowing the space on which the fields are defined to be fundamentally quantum. This approach is at the crossroads of three lines of investigations. First, it stems from the attempt to reconcile the presence of a fundamental length scale with the principle of relativity \cite{AmelinoCamelia:2011bm}. This has led to the idea that locality itself is relative to the space of quantum probes, especially their energy. It also follows from a foundational investigation of string theory that aims to understand what notion of locality is compatible with the symmetry called T-duality,
which exchanges ultraviolet and infrared physics, or more precisely changes what is meant by spacetime. 

Finally, it follows from a deeper study of the geometry of quantum theory that has uncovered new geometrical structures associated with non-commutative algebra, and choices of polarization ($\eta_{AB}$). The central idea is that the same geometric structures ($\omega_{AB}$, $\eta_{AB}$, $H_{AB}$) appear in the study of observer-dependent locality and the geometry of string theory. This strongly suggests that the reconciliation of the quantum and gravity, consistent with the presence of a relativistically invariant fundamental length scale $\lambda$, necessitates the structure constants that define the quantum algebra to be dynamical, on par with the spacetime metric. In other words, the quantum should be ``gravitized,'' and the work of \cite{flm} proposes a precise definition of what this should mean. In particular, a  fundamental geometrical structure, modular spacetime, appears as the effective geometry of compact
strings. This new geometric structure generalizes the usual concept of spacetime and it also embodies the principle of relative locality (that is, observer-dependent locality) by incorporating covariantly a fundamental scale $\lambda$. It can be understood as a new notion of locality that results from the quantum superpositions of spacetimes. On the technical side, gravitizing the quantum implies understanding how to curve modular space consistently, and thus render the above geometric structures,
 $\omega_{AB}$, $\eta_{AB}$, $H_{AB}$, fully dynamical.

{\bf Intrinsic non-commutativity of quantum gravity:} 
One of the central features of \cite{flm} is that one should
work with ``doubled'' albeit non-commutative ``spacetime coordinates'' $(\hat{x}^a, \hat{\tilde x}_b)$
with the following fundamental commutators \cite{flm}\footnote{Note that the sum and difference of these ``doubled'' coordinates can be precisely related to the zero modes of the left and right moving modes of the string \cite{flm}.}
\bea  
[\hat{x}^a,\hat{x}^b]=0,\qquad 
[\hat{x}^a,\hat{\tilde x}_b]=2\pi i\lambda^2 \delta^a{}_b,\qquad 
[\hat{\tilde x}_a,\hat{\tilde x}_b]= 0
\eea
where $\lambda$ denotes the fundamental length scale, such as the Planck scale  $l_P $
(and where  $\lambda \epsilon = \hbar$, $\epsilon$ being the corresponding fundamental energy scale).
Note that full covariance is maintained in this description \cite{flm}, and that the string tension $\alpha'$ is 
given by the ratio of the fundamental length and energy scales $\alpha' = \lambda/{\epsilon}$.
Also, this fundamental non-locality is derived, and not postulated in this new approach to quantum gravity \cite{flm}.
Note that this intrinsic non-commutativity in quantum gravity is directly tied to the above symplectic structure $\omega_{AB}$.

It is important to discuss the question of the physical meaning of $(\hat{x}^a, \hat{\tilde x}_b)$.
By applying the usual logic that connects the non-trivial commutators to their uncertainties, we can understand the 
meaning of $(\hat{x}^a, \hat{\tilde x}_b)$, as follows: the $\hat{x}^a$ coordinate can be associated with short-distance (UV) ``spacetime'' and the $\hat{\tilde x}_b$ with the long-stance (IR) ``spacetime,'' such that their corresponding uncertainties are necessarily complementary, i.e.: $ \Delta {x}^a \Delta {\tilde x}_b \sim \lambda^2 \delta^a_b$.

In what follows, we concentrate on the observational effects of this fundamental non-locality.
That such non-local physics might be a generic feature of quantum gravity, especially in the context of black hole horizon physics, has been recently placed in sharp focus by Joe Polchinski and collaborators \cite{Almheiri:2012rt}.
Therefore we focus on the application of the above fundamental non-commutativity in the context of black hole physics. 
In particular, in view of the string-black hole correspondence, as emphasized by Horowitz and Polchinski \cite{Horowitz:1996nw}, we recall that in order for the typical state of the string to look like the typical state of the black hole, then
$\alpha' \sim r_0^2$, where $r_0$ is the size of the black hole horizon.
This is then the precise context for our exploration of possible observational consequences of the fundamental non-commutativity in quantum gravity.

First we note that according to this new view of quantum gravity local effective fields $\phi(x)$ should be replaced with bi-local fields $\phi(x, \tilde x)$, with $[\hat{x}^a,\hat{\tilde x}_b]=2\pi i\lambda^2 \delta^a{}_b$ \cite{flm}. Such non-commutative field theories \cite{nc} display mixing between
the UV and IR physics, and in order to define such theories in the continuum one has to appeal to a double-scale renormalization group (RG) and the self-dual fixed points. The effective low energy scale can be set as the geometric mean of the naive UV and IR scales, and so the standard claim about the invisibility of quantum gravitational effects at long scales becomes invalid. Let us denote such an effective low energy scale by $L$.

Then one sharp proposal for
a new observable effect based on this intrinsic non-commutativity of spacetime in quantum gravity is:
There should exist observations, in the context of black-hole physics, that correlate
one event at low energy and one event at high energy, in such a way that the product
of their energies/momenta is the square of some observable intermediate energy/momentum scale. (One could roughly think
of these events as EPR-like pairs in energy/momentum space, except that these two ``entangled'' events
should not be originating from a one ``single parent'' event.)

However, such correlated events might be hard to observe, and so we turn to an observable effect directly related
to an effective low energy set by $L$ which could lead to enhanced black-hole mass loss in a BH-NS binary, with fascinating consequences we have already addressed, in a different context in \cite{Simonetti:2010mk}.
Note that there is a formal similarity with the situation discussed in \cite{Simonetti:2010mk}, given the existence of 
new ``shadow'' degrees of freedom associated with the second label $\tilde x$ in the effective bi-local field description of the physics at long distances, $\phi(x, \tilde x)$.

{\bf BH-NS probes of fundamental non-locality:}
As in our original work \cite{Simonetti:2010mk} we consider a binary system consisting of a black hole (BH) and a neutron star (NS), where the NS is a pulsar. As pointed out in that
work, observations of the pulsar could be used to measure the changing orbital period of the system with sufficiently high precision to either directly observe the outspiral behavior, or measure the
competing contributions of mass loss and gravitational radiation as they affect the inspiral rate
of the system. Alternatively, improved limits could be set on the size of the effective scale $L$.
The primary historical motivation here comes from the PSR B1913+16 Binary Pulsar case, where observations of the one NS that acts as a pulsar have yielded high precision
determinations of the parameters of the system, and have provided a dramatic test of relativistic physics \cite{Weisberg:2010zz}.

The BH evaporation rate in the scenario considered in \cite{Simonetti:2010mk} is given by
\be
{\dot{M}}_{BH} =  -2.8\times10^{-7} { M_{BH}^{-2}} L^2\  M_\odot {\rm y}^{-1}
\ee
where $M_{BH}$ is the mass of the black hole in solar mass units ($M_\odot$), and $L$ is some effective length scale coming from the above intrinsic non-commutativity of quantum gravity, in units of 10$\mu$m, that we can take as a phenomenological parameter\footnote{In the context of braneworld scenarios \cite{bw} discussed in \cite{Simonetti:2010mk}, this length scale $L$ was set by the AdS radius in units of 10 $\mu$m.}. Given the presence of a large number of ``shadow'' degrees of freedom, such evaporation can be
enhanced.

Then, following \cite{Simonetti:2010mk}, one can explicitly evaluate the effects of mass loss due to enhanced BH evaporation
and compare them to the effects of energy loss due to gravitational radiation.
It turns out, a shown in \cite{Simonetti:2010mk}, that in a BH-NS system gravitational radiation and BH mass-loss can produce opposite effects leading to, for some masses and orbital parameters in the expected ranges, to the outspiral of the binary components!
This would be a dramatic effect illustrating our main point: the BH-NS binary is a precision probe for quantum gravity effects at large scales. For $L\sim 10~\mu$m, the competition between energy loss leading to inspiral, and black hole mass loss leading to outspiral, can be discerned for observations of the precision used in the PSR B1913$+$16 Binary Pulsar case \cite{Simonetti:2010mk}. For any $L$ larger than 10~$\mu$m, the observable effects will be easier to detect. This implies that the model discussed here can be readily tested.

{\bf Outlook:}
While a BH-NS candidate has not yet been found there is great hope that such a system will be discovered. However, given the existing spectacular results obtained by the LIGO collaboration, it is not unreasonable to expect the observation of the very first BH-NS binary in the 
near future\footnote{A recent reference on this subject is \cite{Yang:2017gfb}.}. Such an observation would necessarily be of a merger event while the type of observation described above will require a stable binary pair. The advent of new instruments like the Square Kilometer Array (SKA) and the Laser Interferometer Space Antenna (LISA) open possibility of the discovery of a stable BH-NS system \cite{Carilli:2004nx, AmaroSeoane:2012km}. Then, as argued in this essay, such a BH-NS binary might provide us with a first window into the physics of quantum gravity.

\vskip 0.5cm

\noindent
{\bf Acknowledgments:} 
We are very grateful to Laurent Freidel and Rob Leigh for numerous insightful discussions over many years on the topic of
quantum gravity. We thank L. Lehner for informative conversations. The work of DM is supported in part by the Julian Schwinger Foundation.

\end{document}